\documentclass{iau} 
\usepackage{graphicx}

\title[Time-domain Astronomy with the GMRT] 
{Time-domain Astronomy with the GMRT: \\ uGMRT to eGMRT}

\author[Jayanta Roy]   
{Jayanta Roy$^1$}

\affiliation{$^1$NCRA-TIFR, Pune University Campus, \\ Pune 411007,
India \\ email: {\tt jroy@ncra.tifr.res.in}} 

\pubyear{2017}
\volume{337}  
\jname{Pulsar Astrophysics -­ The Next 50 Years}
\editors{P. Weltevrede, B.B.P. Perera, L. Levin Preston \& S. Sanidas, eds.}

\begin{document}

\maketitle

\begin{abstract}
The upgraded GMRT (uGMRT) with its unprecedented sensitivity and high figure of merit, is expected to result in the discovery of a 
large population of pulsars including pulsars of previously unknown type. In the phase-2 of the 
GMRT High Resolution Southern Sky (GHRSS) survey with the uGMRT we will reach 1/4th of sensitivity of the SKA Phase-1.
 In this paper we highlight the salient features of the survey 
of pulsars and fast transients with the uGMRT highlighting its discovery potential. The extended GMRT (eGMRT) equipped with wide 
field-of-view, increased collecting area will have unprecedented sensitivity in time-domain astronomy, reaching close to SKA-Phase1. 
Many fold increase in the number of elements, 
increase in the baseline length and addition of phased array feed will make eGMRT an excellent instrument for the survey of pulsars 
and transients with a promise of detecting large variety of pulsars and fast radio bursts. 
\end{abstract}

\firstsection 
\section{Introduction}
Inspite of persistent efforts for last five decades, present known population of pulsar is only 5\% 
or less of the total Galactic population of pulsars (estimates of Galactic population ranges from 
40,000 to 90,000 objects, e.g. \cite[Lorimer 2008]{Lorimer08}), indicating that a vast majority of pulsars 
are waiting to be discovered. In addition to the regular radio emission
from pulsars, millisecond transient bursts, called Fast Radio Bursts (FRBs; \cite[Lorimer et al. 2007]{Lorimer07}, 
\cite[Thornton et al. 2013]{Thorton13}) 
are observed at the locations of dynamic events, making them useful probes for extreme matter states. 
Possible extragalactic origin allows these bursts to be used for determination of baryon content of the 
intergalactic medium.

Because of steep spectral nature of the pulsars, lower frequencies are better choice for
searching the fainter pulsars away from the Galactic plane. Beside the GBT and the
LOFAR, the Giant Metrewave Radio Telescope (GMRT) is another telescope having sensitive
low frequency ($<$ 600 MHz) observing facility. In addition, the lower frequency surveys are also 
benefited from larger field-of-views. Motivated by these and with the aid of 
high time-frequency resolution, reduced quantised noise software backend (GSB; \cite[Roy et al. 2010]{Roy10}), 
we started GMRT High Resolution Southern Sky (GHRSS; \cite[Bhattacharyya et al. 2016]{Bhattacharyya16}) survey 
for pulsars and transients at 322 MHz for a target sky which is not being searched at frequencies below 1.4 GHz for last two decades. 
In phase-1 of the survey completing 1800 deg$^2$ sky coverage, we have discovered 13 new pulsars 
(\cite[Bhattacharyya 2017]{Bhattacharyya17}), which is a very encouraging pulsar-per-deg$^2$ discovery rate.

GMRT is currently undergoing a major upgrade to increase instantaneous bandwidth as well as to improve the receiver systems for better 
$G/T_{sys}$. For example, the fractional bandwidth for 322 MHz system is increased by 5-times resulting in a 2$-$3-times increase 
of the sensitivity. This new upgrade (called uGMRT; \cite[Gupta et al. 2017]{Gupta17}) brings a quantum leap in the survey 
sensitivity for pulsars and FRBs at low and mid frequencies. In Figure 1, a full synthesis continuum sensitivity of the 
GMRT (along with uGMRT) is compared with the other telescopes in operation/building/design. Two dashed lines at a slope of 
$-$0.7 (typical spectral index for extra-galactic sources) represent the VLA and JVLA sensitivities indicating that low and mid 
frequency sensitivity of the uGMRT are comparable to the high frequency sensitivity of the JVLA. Moreover, uGMRT is one of the 
most sensitive facility at these frequencies in the pre-SKA era.    

\begin{figure}[b]
\begin{center}
 \includegraphics[width=3.2in]{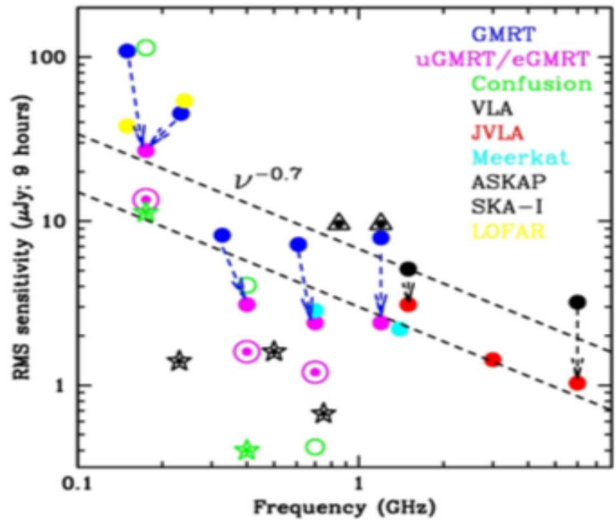}
 \caption{A full synthesis continuum sensitivity of the GMRT (along with uGMRT, marked in solid circle at the tip of the arrows) 
in comparison with the other telescopes 
in operation/building/designing. This illustrates that uGMRT is one of the most sensitive telescope up to 1 GHz till SKA-Phase1 
(marked in asterisks). Figure courtesy: Nissim Kanekar}
   \label{fig1}
\end{center}
\end{figure}

\section{Pulsar survey with uGMRT: GHRSS Phase-2}

Pulsar surveys are sensitivity limited, hence the design of more sensitive instruments promises higher
discovery rate. With the aid uGMRT, in the GHRSS Phase-2 (300 $-$ 500 MHz) at least 2-times improvement 
in sensitivity compared to Phase-1 is expected to be achieved. This is important as we start to probe the lower range of the
pulsar luminosity distribution. Understanding the shape of this distribution will be vital for understanding
how many pulsars the SKA will find. Figure 2 compares estimated theoretical 
sensitivity of the GHRSS survey Phase-2 (with 200 MHz bandwidth) with Phase-1 (with 32 MHz bandwidth) and other surveys.
The sensitivity estimations 
are at 322 MHz for 5$\sigma$ detection with 10\% duty cycle in cold sky. The flux densities are scaled using 
$-$1.4 spectral index to derive the corresponding 322 MHz values. This illustrates that GHRSS survey is the 
most sensitive survey in the GBNCC complementary sky. For the Phase-2 of the GHRSS survey we calculate a 
theoretical survey sensitivity of $\sim$ 0.25 mJy at 400 MHz (considering the radiometer equation) for a 5$\sigma$ 
detection for 10\% duty cycle, with the GMRT incoherent array of gain 2.5 K/Jy for 
200 MHz bandwidth and considering a system temperature of 106 K. 
We also calculate a sensitivity of 0.5 Jy for a 10$\sigma$ detection limit for 
5 ms transient millisecond bursts, considering weak scattering (\cite[Thornton et al. 2013]{Thornton13}). The survey 
parameters for GHRSS Phase-2 and predicted discoveries are presented in Table \ref{tab1}.

\begin{figure}[b]
\begin{center}
 \includegraphics[width=2.8in,angle=-90]{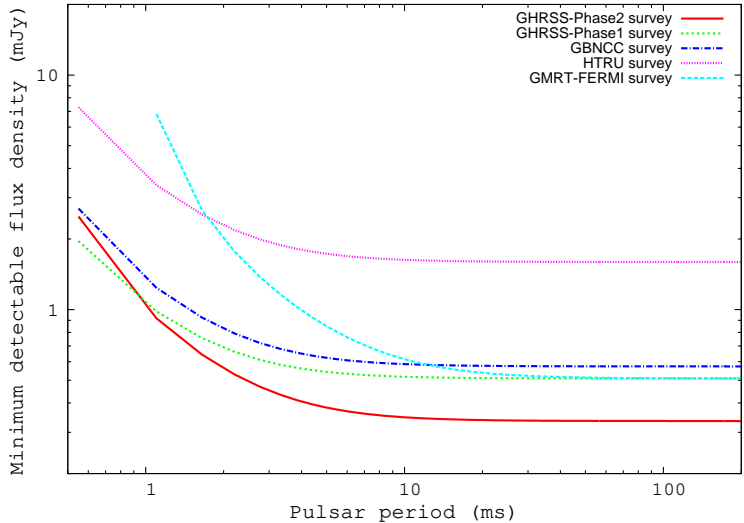} 
 \caption{Estimated sensitivity of the GHRSS survey Phase 1 (green dotted line), Phase 2(with uGMRT; red solid line), 
Fermi-directed search with the GMRT (cyan dashed line), 
GBNCC (blue dashed-dot line) and the HTRU survey (magenta dotted line).}
   \label{fig2}
\end{center}
\end{figure}

\begin{table}
\begin{center}
\caption{Survey parameters of the GHRSS Phase-2 survey}
\label{tab1}
{\scriptsize
\begin{tabular}{|l|c|}
\hline
Galactic region      &  $|b|$ $>$ 5deg \\
Declination          &  Dec$>$$-$ 54deg, Dec$<$$-$ 40deg \\
Integration time     &  600 s            \\
Sampling time        &  81 $\mu$s      \\
Bandwidth            &  200 MHz      \\
Number of channels   &  4096          \\
Frequency Resolution &  48 kHz      \\
Data rate            &  50 MB/s \\
Pulsar Sensitivity   &  0.2 mJy at 5$\sigma$              \\
Single pulse Sensitivity & 0.5 Jy at 10$\sigma$ for 5 ms \\
Single pulse fluence &  1 Jy-ms \\
Data/pointing        &  30 GB  \\
survey area          &  1500 deg$^{2}$ \\
Predicted discovery$^1$ pulsar & 40 PSRs and 10 MSPs \\
Predicted discovery$^2$ FRB & 4$^{+3}_{-2}$ \\ \hline
\end{tabular}
}
\end{center}
\vspace{1mm}
 \scriptsize{
 {\it Notes:}\\
$^1$  PsrPopPy (\cite[Bates et al. 2014]{Bates14})\\
$^2$  \cite[Champion et al. (2016)]{Champion16} \\
}
\end{table}

\begin{figure}[b]
\begin{center}
 \includegraphics[width=5.8in]{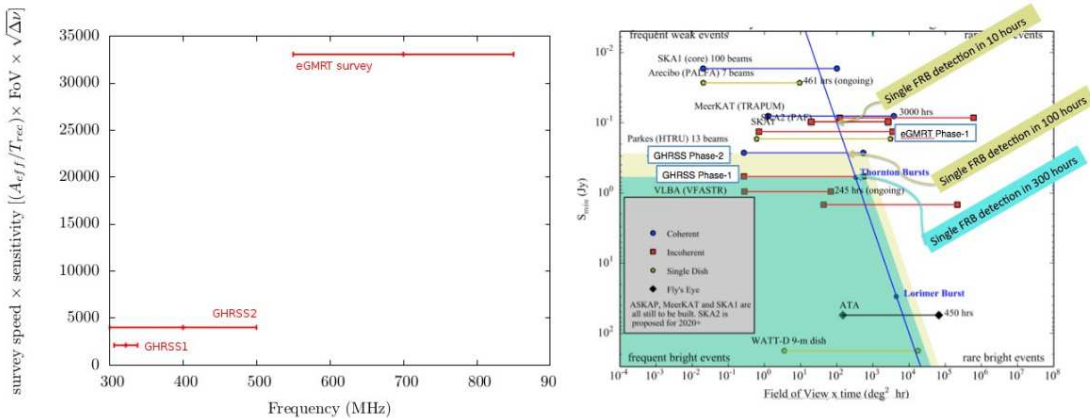}
 \caption{Prospect of time-domain survey with GMRT. Left panel is pulsar survey metric for GHRSS survey Phase-1, 
Phase-2 and eGMRT survey. Right panel is FRB detection capability for GHRSS Phase-1, Phase-2 and eGMRT survey. 
Original figure courtesy: T. Colegate.}
   \label{fig3}
\end{center}
\end{figure}


\section{The eGMRT}
The uGMRT already brings seamless frequency coverage with 2$-$3 times improvements in sensitivity. In order to increase 
the collecting area of the GMRT at least by factor of 2, the design study for the expansion of GMRT (called eGMRT; 
\cite[Patra et al. in preparation]{Patra17}) are now being carried out. The eGMRT with 30 new antennas in 
short and mid baselines along with 30 existing antennas, gives an unprecedented sensitivity at low-frequency comparable to 
SKA Phase-1. 
This can be seen in Figure 1, where the predicted sensitivities for the eGMRT are marked in magenta encircling dots. The eGMRT 
equipped with 30 beams phased array feed (PAF) for each antenna, resulting in an order of magnitude increase of the field-of-view, 
can be an excellent survey instrument for pulsars and transients. A survey metric for pulsars moving from GHRSS Phase-1 to 
GHRSS Phase-2 with uGMRT to eGMRT Phase-1 (possibly with 60 antennas each with PAF at 550$-$850 MHz band) are showing in Figure 3 (left panel). 
Survey speed weighted by instantaneous sensitivity experiences a factor of 2 improvement in the GHRSS Phase-2, which will 
be enhanced by 7$-$8 times with the eGMRT. The prospect of detecting FRBs are illustrated in Figure 3 (right panel). The occupancy 
in single-pulse search space for GHRSS Phase-1 and Phase-2 are marked by green and yellow regions. GHRSS Phase-1 sensitivity is 
crossing the single FRB detection threshold (guided by Lorimer and Thorton bursts ignoring frequency dependent scattering and 
spectral steepening) in 300 hours of observing time, which reduces to 100 hours for Phase-2. For survey with eGMRT Phase-1, in 
every 10 hours, one FRB could possibly be detected. Thus, moving from GHRSS Phase-1 to eGMRT, steeping through GHRSS Phase-2 with 
uGMRT, provides an excellent pathways to bring renaissance in the field of time-domain processing.

\end{document}